\documentstyle[12pt]{article}

\newcommand{\Eins}{1\hspace{-0.12cm}\mbox{I}}

\begin{document}

\title{Functional Integral Approach to the single impurity
	Anderson Model}

\author{R.~Bulla, J.~Keller, T.~Pruschke \\ Institut f\"ur Theoretische
Physik, Universit\"at Regensburg \\93040 Regensburg, Germany}

\maketitle

\begin{abstract}

Recently, a functional integral representation was proposed by Weller
\cite{Wel90a}, in which
the fermionic fields strictly satisfy the constraint of no double occupancy at
each lattice site. This is achieved by introducing spin dependent Bose fields.
The functional integral method is applied to the
single impurity Anderson model both in the Kondo and mixed-valence regime.
The f-electron Green's function and susceptibility are calculated using
an Ising-like representation for the Bose fields.
We discuss the difficulty to extract a spectral function from the
knowledge of the imaginary time Green's function.
The results are compared with NCA calculations.

keywords: functional integral, single impurity Anderson model,
f-electron Green's function
\end{abstract}

\section{Introduction}

In systems of correlated fermions on a lattice one often encounters the
presence
of a very high local Coulomb repulsion $U$ between two particles on the same
lattice site. In this case simple perturbation theory in the parameter $U$
no longer provides a good approximation.
In the limit $U=\infty$, the interaction term in the
Hamiltonian can be eliminated completely if a constraint is introduced
which allows only empty and singly occupied states.

Functional integral techniques are a very powerful tool for the investigation
of lattice fermion systems and there are several ways to incorporate the
constraint
in the functional integral. Starting for example with a slave boson approach,
the constraint is guaranteed by delta-functions in the integration measure.

The method we consider here is based on the idea of projecting out the doubly
occupied sites already in the derivation of a coherent state functional
integral. The new theory was proposed and studied by Weller \cite{Wel90a} and
has already been applied to the
one-\cite{Wel90b} and two-\cite{Zha91a,Zha91b} dimensional Hubbard model.

In this paper, we focus on the single impurity Anderson model (SIAM) with
infinite local Coulomb repulsion at the impurity site. This model was proposed
by Anderson \cite{And61} to describe the properties of metals containing
magnetic
impurities and has been intensively studied
(see e.g.~\cite{Bic87,Tsv83,Kri80}), so that good approximate results
are available today. However, these methods do not work for all
parameter ranges of the model or cannot easily be extended to the lattice
case. Therefore
there is still some need for alternative methods to solve the impurity Anderson
model.

We have to state here that also with our method, we do not succeed in
obtaining more satisfying results for the SIAM. Nevertheless, we think
it is worth publishing the results because the new method is conceptionally
quite different to the approaches cited above and on the other hand may be
useful in the future if the limitations by computer time become less serious.
Even analytical approaches based on
this functional integral method may become possible, perhaps also in the
investigation
of other models of strongly correlated electron systems.

The rest of the paper is organized as follows. After an introduction to the
new functional integral method (section 2) we derive an effective action
for the impurity states of the SIAM by
integrating over the conduction electron degrees of freedom (section 3). This
action is used as the starting point for the numerical investigation described
in
section 4. In section 5 we present results for the imaginary time Green's
function $G(\tau)$ from which the f-occupancy can be derived immediately.
Using a fit
procedure we obtain the spectral function $A(\omega)$. To demonstrate
the limitation of the fit procedure we show that various quite different
spectral functions can result in very excellent fits to the same given
$G(\tau)$.
We also discuss the case of the impurity f-level lying at the Fermi level.
Here we expect the spectral function to
consist of a single Lorentzian. Within this restriction we
get a broadening and renormalization of the f-level consistent with
calculations within a
second order perturbation theory in the hybridization. We compare the
results with NCA (Non Crossing Approximation) calculations for both
parameter sets.
With an additional magnetic field coupling to the impurity state
we calculate the static susceptibility and the effective magnetic moment.

\section{Theory}

In this section we follow closely the method originally published
by Weller \cite{Wel90a}. We want to introduce the functional integral method
by considering first the simple
example of a fermionic system with one site and infinite Coulomb repulsion
\begin{equation}
   H = \sum_{\sigma}\varepsilon_{\rm f} f^\dagger_{\sigma}f_{\sigma}
     + U f^\dagger_\uparrow f_\uparrow f^\dagger_\downarrow f_\downarrow
      \ \ \ \ (U \to\infty) \ \ \ \ .
\end{equation}
Introducing Hubbard operators
\begin{equation}
X_{0\sigma} = \vert\, 0\rangle\langle\sigma\vert \ \ \ \ \ {\rm and}  \ \ \ \ \
X_{\sigma 0} = \vert\, \sigma\rangle\langle 0\vert
\end{equation}
 which project
out the doubly occupied states, the Hamiltonian can then be written as
\begin{equation}
   H = \sum_{\sigma}\varepsilon_{\rm f} X_{\sigma 0} X_{0\sigma} \ \ \ \ .
\end{equation}

Essential for the calculation of statistical properties of the fermionic
system at finite temperatures is the partition function
\begin{equation}
Z=Tr\left[e^{-\beta \hat H} \right] \ \ \ \ \ \big(\beta = {1\over
   {k_{\rm B} T}} \big)
\end{equation}
with the trace in the restricted Hilbert space. We divide
$\exp(-\beta \hat H)$ into a product of $N$ equal operators
$\exp(-\Delta\tau \hat H)$, thereby defining steps $\tau_n=(n-1)\Delta\tau$
$(n=1,\ldots N,(\Delta\tau =
      {\beta / N})  )$
in the imaginary time interval $[0,\beta]$. At each time step we
insert unity operators $\Eins_n$ which
project out the doubly occupied states.
\begin{eqnarray}
 Z &=& Tr\left[e^{-\Delta\tau H} e^{-\Delta\tau H}
\dots e^{-\Delta\tau H}\right] \\
 &=&Tr\left[e^{-\Delta\tau H} \Eins_N e^{-\Delta\tau H}
 \Eins_{N-1}
\dots \Eins_2 e^{-\Delta\tau H} \right]
\end{eqnarray}
with
\begin{equation}
\Eins_n =\vert\, 0\rangle\langle 0\vert+\vert \uparrow
\rangle\langle\uparrow\vert +
\vert \downarrow \rangle\langle\downarrow\vert \ \ \ .
\end{equation}
The unity operators actually do not depend on $n$. This index is used to
distinguish auxiliary fields which will be introduced at each time step $n$
(see (8)).
The most elegant way to derive a functional integral for fermionic systems
is to use projectors
expressed in terms of coherent states (see e.g. \cite{Neg88}).
Following Weller \cite{Wel90a}, we make the ansatz:
\begin{equation}
\Eins_n = \int_{b_n}\int {\rm d}\psi_n^\dagger {\rm d}\psi_n
	\vert\, \psi_n,b_n\rangle\langle
	\psi_n,b_n\vert
\end{equation}
with the coherent states
\begin{equation}
  \left\vert\, \psi_n,b_n\right\rangle = \left [1+{1 \over 2}
 \psi_n\psi_n^\dagger
  +\sum_\sigma \psi_n b_{\sigma n} X_{\sigma 0} \right]\left\vert\, 0
  \right\rangle  = \exp\left [{1 \over 2} \psi_n\psi_n^\dagger
  +\sum_\sigma \psi_n b_{\sigma n} X_{\sigma 0} \right]\left\vert\, 0
  \right\rangle
\end{equation}
which are constructed to be eigenstates of the operator $X_{0\sigma }$
\begin{equation}
X_{0\sigma } \left\vert\, \psi_n,b_n\right\rangle = \psi_nb_{n\sigma}
\left\vert\, \psi_n,b_n\right\rangle \ \ \ .
\end{equation}
The $\psi$ are Grassmann variables obeying the usual anticommutation relations.
The $b$-variables are ordinary complex numbers and do not have a direct
physical meaning. They just count the spin
multiplicity for each step in the imaginary time interval.
Because of the restriction to zero or singly occupied sites and because
we used spin carrying $b$-variables in the ansatz (9), we only need {\it one}
Grassmann field for each time step $n$.

{}From requiring the equality of equations (6) and (7),
it is straightforward to prove that the following equations have to be
fulfilled:
\begin{equation}
  \int_b b_\uparrow^* b_\uparrow =\int_b b_\downarrow^* b_\downarrow =1
  \ ;\ \ \ \
  \int_b b_\uparrow^* b_\downarrow =\int_b b_\downarrow^* b_\uparrow =0
  \ ;\ \ \ \
  \int_b 1 = 1
 \end{equation}
In \cite{Wel90a} several possible representations for the $b$-variables
are presented. Below we will use a special spin representation (see (30)).

Replacing the trace by
\begin{equation}
  Tr[\dots] \to \int_{b_1}\int {\rm d}\psi_1^\dagger {\rm d}\psi_1
	\langle\psi_1,b_1\vert\dots\vert\, \psi_1,b_1\rangle\ \ \ ,
\end{equation}
the next step is to evaluate the matrix elements
$\langle\psi_n,b_n\vert e^{-\Delta\tau H}
\vert\, \psi_{n-1},b_{n-1}\rangle  $. For nontrivial Hamiltonians this can only
be
performed in the limit
$\Delta\tau \to 0$, and therefore the calculations become
exact only in the limit $N\to\infty$.

Finally we obtain the functional integral representation for the partition
function:
\begin{equation}
Z = \int_{b}\int {\cal D}\psi^\dagger {\cal D}\psi\
\exp\left[ {S(\psi,\psi^\dagger,b)}  \right]
\end{equation}
with:
\begin{equation}
S(\psi,\psi^\dagger,b) = \sum_{n=1}^N \psi^\dagger_n
  \left( \psi_{n-1} \left( \sum_\sigma b^*_{\sigma n} b_{\sigma n-1} \right)
    - \psi_n \right)  $$
    $$ -\Delta\tau \sum_{n=1}^N H\left(X_{\sigma 0}\to \psi^\dagger_n
    b^*_{\sigma n},X_{0\sigma } \to \psi_{n-1}
    b_{\sigma n-1} \right)
\end{equation}
and
\begin{equation}
    \int {\cal D}\psi^\dagger {\cal D}\psi\ = \int {\rm d}\psi^\dagger_N
{\rm d}\psi_N
    {\rm d}\psi^\dagger_{N-1} {\rm d}\psi_{N-1} \ldots {\rm d}\psi^\dagger_1
    {\rm d}\psi_1
\end{equation}
Note that --- due to the $\sum_\sigma b^*_{\sigma n} b_{\sigma n-1}$
term --- we do not arrive at the usual expression with the time derivative
($\psi^\dagger {\partial \over {\partial\tau}} \psi$) in the
kinetic part of the action!

The functional integral representation for the imaginary time Green's function
\begin{equation}
  -G(\tau) = \left\langle X_{\sigma 0}(\tau)X_{0\sigma } (0) \right\rangle
  = {1 \over Z} Tr\left[e^{-\beta H}
  X_{\sigma 0}(\tau)X_{0\sigma } (0) \right]
\end{equation}
can be derived in a quite similar way as described for the partition function.
The result is
\begin{equation}
  -G(\tau)  = {1 \over Z}\int_{b}\int {\cal D}\psi^\dagger {\cal D}\psi\
       \psi_1 \psi_n^\dagger b_{\sigma 1} b^*_{\sigma n}
       \exp\left[ {S(\psi,\psi^\dagger,b)} \right ]
\end{equation}
with
\begin{equation}
\tau = (n-1) {\beta \over N} \ \ ,\ \ n=1\dots N\ \ \ \  .
\end{equation}
Note that in the integrand Grassmann variables always appear in products with
$b$-variables.

\section{Action for the ($U=\infty$) Single Impurity Anderson Model}

Now we want to derive an effective action for the f-electrons
of the SIAM by integrating out the conduction electron degrees of freedom.
This action will be the  starting point for the numerical investigation
described in section 4.

In the limit $U\to\infty$ the Hamiltonian of the SIAM is given by:
\begin{eqnarray}
H&=& \sum_{\sigma}\varepsilon_{\rm f} X_{\sigma 0} X_{0\sigma}
	+ \sum_{k\sigma } \varepsilon_k c^\dagger_{k\sigma } c_{k\sigma }
     \nonumber  \\
  & & + \sum_{k\sigma } V \left(   X_{\sigma 0} c_{k\sigma }
			 + c^\dagger_{k\sigma }  X_{0 \sigma} \right)
\end{eqnarray}
The operators $c^\dagger_{k\sigma }$ ($ c_{k\sigma }$) create (annihilate)
a conduction electron with spin $\sigma$ and wavevector ${\bf k}$.
The last term describes the hybridization between the localized f-state and
the conduction electrons with the hybridization matrix element
$V$ independent of ${\bf k}$.
The energies $\varepsilon_k$ and
$\varepsilon_{\rm{f}}$ are to be measured from the Fermi level.

The action corresponding to the above Hamiltonian is:
\begin{eqnarray}
S(\psi,\psi^\dagger,\chi,\chi^\dagger,b) &=& \sum_{n=1}^N \psi^\dagger_n
  \left( (1-{\beta \over N}\varepsilon_{\rm f})
  \psi_{n-1} \bigl( \sum_\sigma b^*_{\sigma n} b_{\sigma n-1} \bigr)
    - \psi_n \right) \nonumber \\
 &+& \sum_{nk\sigma} \chi^\dagger_{k\sigma n}
  \left( (1-{\beta \over N}\varepsilon_k)
  \chi_{k\sigma n-1}  - \chi_{k\sigma n} \right) \nonumber\\
   &-& {{\beta V} \over N} \sum_{nk\sigma} \left(
     \psi^\dagger_n b^*_{\sigma n} \chi_{k\sigma n-1}
    + \chi^\dagger_{k\sigma n}  \psi_{n-1}b_{\sigma n-1} \right)
\end{eqnarray}
For the conduction electron operators $c^\dagger_{k\sigma }$ and
$ c_{k\sigma }$ (whose corresponding Hilbert space is not restricted)
we applied the standard method of replacing
operators by Grassmann variables (see e.g.~\cite{Neg88}).
Now we have to integrate over three types of fields:
the complex $b$-fields, the Grassmann variables $\psi$ for the
f-electrons and the Grassmann variables $\chi$ for the conduction electrons.
The partition function for the whole system then reads:
\begin{equation}
Z = \int_{b}\int {\cal D}\psi^\dagger {\cal D}\psi\
       \int {\cal D}\chi^\dagger {\cal D}\chi\ \exp\left[
       {S(\psi,\psi^\dagger,\chi,\chi^\dagger,b)} \right]
\end{equation}
The integrations can in principle be performed in several ways;
we choose to first integrate over the Grassmann fields of the conduction
electrons using the formula:
\begin{equation}
  \int {\cal D}\chi^\dagger {\cal D}\chi \exp\left[{-\sum_{ij}\chi^\dagger_i
H_{ij}
  \chi_j +\sum_i \chi_i \xi_i^\dagger +\sum_i \chi_i^\dagger \xi_i}
  \right] =  (\det  H) \exp\left[
  {-\sum_{ij}\xi^\dagger_i  (H^{-1})_{ij}  \xi_j}  \right]
\end{equation}
The resulting action
\begin{equation}
S_{\rm eff}(\psi,\psi^\dagger,b)
  = S_{\rm f}(\psi,\psi^\dagger,b) + \left( {{\beta  V} \over N} \right)^2
\sum_{\sigma  n  m}  \psi^\dagger_{n+1}   b^*_{\sigma  n+1}
\psi_{m-1} b_{\sigma m-1}    \sum_k
(H^{-1})_{n m}(k)
\end{equation}
($S_{\rm f}$ is the first term at the right hand side of (20).)
reduces the problem to that of a single electron with a time dependent
coupling mediated by the conduction electron Green's function $H^{-1}$
with the matrix elements:
\begin{equation}
(H^{-1})_{n m}(k) = \left\{ \begin{array}{r@{\quad:\quad}l}
		     \left(1 + e^{-\beta\varepsilon_k} \right)^{-1}
		     \exp\left({(n-m){\beta \over N} \varepsilon_k}
        \right)& n\ge m \\
		     -\left(1 + e^{-\beta\varepsilon_k} \right)^{-1}
		     \exp\left({(N-(m-n)){\beta \over N} \varepsilon_k}
         \right) & n < m
			     \end{array} \right.
\end{equation}

Using the formula
\begin{equation}
\int {\cal D}\psi^\dagger {\cal D}\psi \exp\left[ {-\sum_{ij}\psi^\dagger_i
M_{ij}
\psi_j} \right]  = \det M
\end{equation}
with
\begin{eqnarray}
 M_{nm}(b) &=& (1-{\beta \over N}\varepsilon_{\rm f})
  \bigl( \sum_\sigma b^*_{\sigma n} b_{\sigma n-1} \bigr) \delta_{n-1,m}
  - \delta_{nm} \nonumber \\
  &+&  \left( {{\beta  V} \over N} \right)^2
\sum_{\sigma}  b^*_{\sigma  n}
b_{\sigma m}    \sum_k
(H^{-1})_{n-1, m+1}(k)
\end{eqnarray}
the integration over the remaining Grassmann variables $\psi_{1\ldots N}$
leads to
\begin{equation}
Z = (\det H) \int_b \det M(b)
\end{equation}
for the partition function. For the imaginary time Green's function we get
\begin{equation}
  -G(\tau)
  = (\det H)
  {1 \over Z} \int_b b_{\sigma 1} b^*_{\sigma n} (M^{-1}(b))_{1,n}
  \det M(b) \ \ \ \ .
\end{equation}
Note that this equation defines $G(\tau)$ only for the $\tau$-values
 $  \tau = (n-1){\beta\over N} \ (n=1,\ldots N)$ .
$G(\beta)$ cannot be calculated directly within this approach but is
simply related to $G(0)$ by
\begin{equation}
      G(\beta) = 1 - 2G(0)
\end{equation}
in the case of no double occupancy.
We do not actually need to
calculate $\det H$ because this factor cancels in the Green's function.

\section{Numerical Investigation}

The resemblance of the theory
to slave boson techniques and the success of slave boson mean field approaches
may suggest that also in our case a saddle-point like approximation for the
$b$-fields is possible. But due to the unusual kinetic part in the action
any replacement of e.g.~$\sum_\sigma b^*_{\sigma n} b_{\sigma n-1} $
by a constant number not equal to one leads to divergences in
the limit $N\to\infty$. So far we did not find any analytical
approach with the action (23) as a starting point and therefore
restrict ourselves to numerical results in this paper.

The most suitable representation of the $b$-fields for a numerical
implementation of the equations (27) and (28) is an Ising-like
representation \cite{Wel90a}:
\begin{equation}
b_{\uparrow n} =1 \ \ \ \ \ \ \ \ \ \ \ \ \ \ b_{\downarrow n} =\pm 1
\end{equation}
\begin{equation}
\int_b \dots = {1\over 2} \sum_{b_{\downarrow n} =\pm 1} \dots
\end{equation}
so that the sums in (27) and (28) contain $2^N$ terms.
For each of these contributions we have to calculate the Matrix $M$
defined by (26). The sum over $k$ in (26) is calculated
using a constant density of states for the conduction electrons in the
interval $[-D,D]$.
\begin{equation}
  \sum_k (H^{-1})_{n m}(k) = \int {\rm d}\varepsilon \rho(\varepsilon)
  (H^{-1})_{n m}(\varepsilon)  = \rho \int_{-D}^D{\rm d}\varepsilon
  (H^{-1})_{n m}(\varepsilon)
\end{equation}
These integrations are independent of the $b$ fields and therefore have to be
performed only once.
NAG routines are used to calculate the inverse and the determinant of
$M$. For $N=20$
the computation and summation of all contributions to $G(\tau)$ takes about
one day CPU time on a workstation.

At this point, one would like to reduce the computer time by restricting the
summation to the most important contributions. This idea fails in our case!
Due to the $\sum_\sigma b^*_{\sigma n} b_{\sigma n-1}$-term, contributions
to the Green's functions can differ by a factor $2^0$ to $2^N$, so that simple
Monte Carlo methods (like the Metropolis algorithm) are unable to explore
the whole phase space of the $b$-variables. On the other hand we experienced,
that
taking into account only the most important contributions, the result is far
away from the exact result received by summing over all contributions.

\section{Results}

Fig.~1 shows the imaginary time Green's function for the parameters
$\varepsilon_{\rm f} = -0.2$,$V=0.22$ and the inverse temperature $\beta =3$
(energies in arbitrary units). It also contains the result of an NCA
calculation which will be discussed below.

The f-occupancy equals
\begin{equation}
   n_{\rm f} =  n_\uparrow + n_\downarrow = 2\cdot G(\tau = \beta) = 0.72 \ \ \
{}.
\end{equation}
Although $G(\tau)$ seems to be rather structureless and indeed has
no direct physical meaning, it is related to the spectral function by
the transformation
\begin{equation}
 -G(\tau) = \int_{-\infty}^\infty {\rm d}\omega e^{-\omega\tau} {A(\omega)
\over
   {1 + e^{-\beta\omega}} } \ \ \ \ \ \ .
\end{equation}
It is easy to calculate $G(\tau)$ with the knowlegde of $A(\omega)$
but {\it not} vice versa. A lot of methods can be found in the literature to
overcome this problem (see e.g.~\cite{Gub91}) but determination of the spectral
function $A(\omega)$ out of $G(\tau)$ remains an extremely ill posed problem as
will
be shown below in an example.

In order to allow the occurrence of an Abrikosov-Suhl resonance in addition
to a broad peak at the f-level position
we tried to fit the numerical
$G(\tau)$ data with a superposition of two Lorentz functions with variable
weight, position and width as the input spectral function. The results are
shown
in Fig.~1 and 2a. Altough we are able to fit the numerical data with an
accuracy
better than $5\cdot 10^{-3}$, several different fits are possible (only three
of the innumerable possibilities are shown). Despite the chance of finding
a spectral function fitting the data even more accurately, we do not think
that it makes sense to decide between different fits, all having this
extreme high accuracy (the $G(\tau)$ data themselves are not exact due
to the finite $N$ and the finite resolution of the computer)!
Therefore we conclude that structures with small weight like an additional
peak at the Fermi level cannot be derived from the knowlegde of $G(\tau)$.

In Fig.~2b the result of a NCA calculation for the same set of parameters is
shown. If the transformation (34) is applied to the spectral function
the dotted line in Fig.~1 is obtained. We observe a significant difference
between the NCA result and our numerical calculation (note the position of
the maximum in the spectral function). Here we cannot decide which theory
gives the better results because both are in a certain sense approximative
(due to the finite $N$ in the functional integral calculation and the leaving
out of crossing diagrams respectively).

Fig.~3a shows results for $G(\tau)$ in case the f-level
is equal to the chemical potential for different values of the hybridization.
For $V=0$ the empty and singly occupied states have equal probabilities:
\begin{equation}
   n_0=n_\uparrow = n_\downarrow = {1 \over 3}
\end{equation}
This is the mixed-valence regime of the SIAM.
With increasing hybridization more and more f-electrons are transferred
to the conduction electron states.
Therefore $n_{\rm f}$ decreases
with increasing $V$ (Fig.~3b). This decrease is a little bit stronger
in the NCA data but proportional to $V^2$ in both cases.

We expect the spectral function to consist of only one Lorentz peak.
With this input knowledge the fit procedure always leads to a well defined
result (if the true spectral function really resembles a single Lorentzian).
Again the accuracy is very high but decreases with increasing hybridization.
The results are shown in Fig.~4a. For zero hybridization we would get
a delta function at $\omega = 0 $. If we increase the hybridization we observe
a broadening and a shift of the peak to higher frequencies. Both effects are
proportional to $V^2$ --- this corresponds to calculations within a second
order perturbation theory.

The NCA spectral functions (Fig.~4b) give qualitatively the same picture. They
are broader and shifted to higher frequencies than the
functional integral results.

The calculation of the f-susceptibility and the corresponding effective
magnetic moment is straightforward within the theoretical scheme described
above. We just add to the Hamiltonian a term
\begin{equation}
   H_{\rm B} = -g\mu_{\rm B} B \left( f^\dagger_\uparrow f_\uparrow
   - f^\dagger_\downarrow f_\downarrow \right)
\end{equation}
and calculate the f-occupancies $n_\uparrow$ and $n_\downarrow$.
The susceptibility is then defined by
\begin{equation}
  \chi = \left. {{\partial (n_\uparrow-n_\downarrow)} \over {\partial B}}
  \right\vert_{B=0}
  \ \ \ .
\end{equation}
In Fig.~5 we show the dependence of the effective magnetic moment
  $\chi / {\beta g \mu_{\rm B}} $
on the temperature for different values of the hybridization.
For zero hybridization, the effective magnetic moment takes its maximum value
at $T=0$. For $V\ge 0.1$, due to the screening of the conduction electrons,
the effective magnetic moment decreases with decreasing temperature.
Of course, the exact behaviour for $T\to 0$ cannot be extrapolated from
the data which are restricted to temperatures higher than $k_{\rm B}T\approx
0.2$ .

\section{Conclusion}

In this paper have we investigated the $U=\infty$ single impurity Anderson
model
using a new functional integral technique in which the constraint of no double
occupancy of the impurity site is fulfilled exactly by auxiliary complex
fields. We obtained an effective action for the impurity as the starting point
for the numerical calculation of the f-electron Green's function.

We discussed the difficulties to extract detailed information
for the spectral function out of the imaginary time Green's function.
In the mixed-valence regime, where we expect that a single peak function is
a good approximation for
the spectral function, we can confirm a shift and broadening of the peak
at the impurity level proportional to $V^2$.

Comparison with NCA results showed small but significant differences.
The origin of these differences is not yet clear.

Calculation of the susceptibility at the impurity showed the expected
temperature dependence. While for zero hybridization, the effective magnetic
moment reaches its maximum value for $T\to 0$, for finite hybridization
we observe a decrease of ${\chi / {\beta g \mu_{\rm B}}} $ with decreasing $T$.

\section{Acknowledgements}

We would like to thank Prof.~W.~Weller for many helpful discussions.
This work has been partly supported by the Graduiertenkolleg ``Komplexit\"at
in Festk\"orpern: Phononen, Elektronen und Strukturen'' at the University
of Regensburg.

\vfill

R.~Bulla, J.~Keller, T.~Pruschke \\ Institut f\"ur Theoretische
Physik\\ Universit\"at Regensburg \\93040 Regensburg\\ Germany \\
email: bulla@rphs1.physik.uni-regensburg.de\\
tel: 0941-943 2046\\
fax: 0941-943 4382

\hfill
\vfill\pagebreak

{\large\bf Figure captions}\\

\vspace{0.5cm}
\begin{itemize}
\item[{\bf Fig.~1}] Green's function $G(\tau)$ on the imaginary time axis
(parameters: $V=0.22, \varepsilon_{\rm f} = -0.2, \beta = 3.0,
D = 3.5, \rho=1, N = 20$ ). The crosses are the numerical result. Also
shown are fits (nearly indistinguishable) calculated from the three
spectral functions in Fig.~2a. The dotted line is the NCA result calculated
from Fig.~2b.
\item[{\bf Fig.~2}] {\bf a)} Three different spectral functions whose
corresponding
$G(\tau)$
(see Fig.~1) fit the numerical data equally well. {\bf b)} Comparison of the
NCA spectral function with fit 1
of Fig.~2a.
\item[{\bf Fig.~3a}] Imaginary time Green's function for $\varepsilon_{\rm f} =
\mu = 0, \beta = 2.0, D=3.5, \rho =1,N=20 $ and different values of the
hybridization.
\item[{\bf Fig.~3b}] Dependence of the f-occupancy $n_{\rm f}$ on the
hybridization (same
parameters as in Fig.~3a).
\item[{\bf Fig.~4}] {\bf a)} Fits for the spectral function to the data shown
in Fig.~3a
with a single Lorentzian. {\bf b)} The NCA spectral
functions for the parameter
set of Fig.~3a.
\item[{\bf Fig.~5}] Dependence of the effective magnetic moment
 on the temperature for different values of the hybridization (parameters:
 $\varepsilon_{\rm f} = -0.2, D=3.5, \rho =1,N=16$).
 The line corresponds to the analytical result for $V=0$.
\end{itemize}

\end{document}